\documentclass[aps,prx,reprint,superscriptaddress]{revtex4-2}

\usepackage{graphicx}
\usepackage{dcolumn}
\usepackage{bm}
\usepackage{amssymb,amstext,amsmath}
\usepackage{xr}
\usepackage{color}
	 \definecolor{darkred}{rgb}{0.75,0,0}
	 \definecolor{darkgreen}{rgb}{0,0.5,0}
	 \definecolor{darkblue}{rgb}{0,0,0.75}
  	 \definecolor{darkorange}{rgb}{1,0.9,0.1}
	 \definecolor{dark}{rgb}{0,0,0}

\begin{document}

\preprint{APS/123-QED}

\title{Temporal networks provide a unifying understanding of the evolution of cooperation}

\author{Aming Li}
\affiliation{%
Center for Systems and Control, College of Engineering, Peking University, Beijing 100871, China}
\affiliation{
Center for Multi-Agent Research, Institute for Artificial Intelligence, Peking University, Beijing 100871, China}
\author{Yao Meng}
\affiliation{%
Center for Systems and Control, College of Engineering, Peking University, Beijing 100871, China}
\author{Lei Zhou}%
\affiliation{%
School of Automation, Beijing Institute of Technology, Beijing 100081, China}
\author{Naoki Masuda}%
\affiliation{%
Department of Mathematics, State University of New York at Buffalo, Buffalo, New York 14260, USA}
\affiliation{%
Computational and Data-Enabled Science and Engineering Program, State University of New York at Buffalo, Buffalo, New York 14260, USA
}%
\author{Long Wang}%
\affiliation{%
Center for Systems and Control, College of Engineering, Peking University, Beijing 100871, China}
\affiliation{
Center for Multi-Agent Research, Institute for Artificial Intelligence, Peking University, Beijing 100871, China}


\date{\today}

\begin{abstract}

Understanding the evolution of cooperation in structured populations represented by networks is a problem of long research interest, and a most fundamental and widespread property of social networks related to cooperation phenomena is that the node's degree (i.e., number of edges connected to the node) is heterogeneously distributed. 
Previous results indicate that static heterogeneous (i.e., degree-heterogeneous) networks promote cooperation in stationarity compared to static regular (i.e., degree-homogeneous) networks if equilibrium dynamics starting from many cooperators and defectors is employed.
However, the above conclusion reverses if we employ non-equilibrium stochastic processes to measure
the fixation probability for cooperation, i.e., the probability that a single cooperator successfully invades a population.
Here we resolve this conundrum by analyzing the fixation of cooperation on temporal (i.e., time-varying) networks. 
We theoretically prove and numerically confirm that on both synthetic and empirical networks, contrary to the case of static networks, temporal heterogeneous networks can promote cooperation more than temporal regular networks in terms of the fixation probability of cooperation. 
Given that the same conclusion is known for the equilibrium fraction of cooperators on temporal networks, the present results provide a unified understanding of the effect of temporal degree heterogeneity on promoting cooperation across two main analytical frameworks, i.e., equilibrium and non-equilibrium ones.

\end{abstract}

\maketitle

\section{Introduction}

The emergence of cooperation through strategy competition and replacement in society of interacting individuals is a crucial phenomenon \cite{Hofbauer98book,Levin2014,HilbeNature2018,hilbe2014cooperation,rapoport1965prisoner,Hauert04Nature,lieberman2005evolutionary,santos2005scale, Traulsen2005,Gomez2007,PercReview17,Szolnoki2012,Zhou2021,Su2019}.
Patterns of interaction among individuals can often be modeled by networks, in which nodes indicate individuals, and edges represent who interacts with whom. Even before we had solid empirical understanding of how humans or animals are connected as networks, Nowak and May showed that the spatial structure of the network, as represented by the square lattice network, for example, promotes cooperation in evolutionary dynamics of the prisoner's dilemma game \cite{Nowak92Nature}. In fact, many real-world contact networks are heterogeneous in the node's degree (i.e., number of edges that the node owns) and are close to scale-free networks, i.e., those with power-law degree distributions \cite{Barabasi1999a,newman2018networks}. It was shown that scale-free networks also promote cooperation in social dilemma situations \cite{santos2005scale,Duran2005}.

\begin{figure}[!h]
	\centering
	\includegraphics[width=0.5\textwidth]{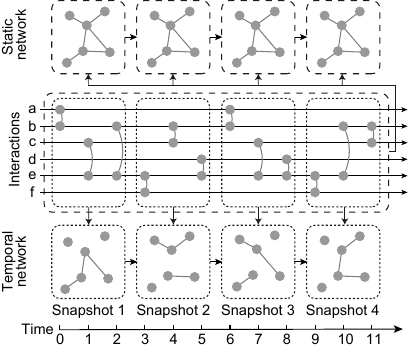}
	\caption{Illustration of temporal and static networks.
		Static networks are constructed by aggregating all interactions (i.e., edges) between each pair of individuals (i.e., nodes) over the entire observation period, and it does not change with time.
		The temporal network is formed by aggregating the time-ordered interactions with a time window of length $\Delta t$ (here $\Delta t=3$), which results in a sequence of network snapshots.
	} \label{Fig1}
\end{figure}

These and many other studies of cooperation using evolutionary game theory, in both well-mixed populations and networks, 
initialize the dynamics with a macroscopic number of cooperators and defectors, such as an equal number of cooperators and defectors, and evaluate the frequency of cooperators in the quasi-equilibrium state, i.e., after a transient time. Beyond this equilibrium dynamics approach, which is important and usually a standard approach when the population is large and its structure is complicated, a drastically different and theoretical approach to the same questions is 
to examine the fixation of cooperation~\cite{Nowak2004,Ohtsuki06Nature,Allen2017}. 
In this non-equilibrium dynamics approach, rooted in population genetics and related probability theory,
one typically starts from a single cooperator, continues a stochastic process of evolutionary dynamics until the fixation (i.e., unanimity) of cooperation or that of defection is reached, and measures, first of all, the probability of fixation of cooperation compared to control cases.
By focusing on the extreme case, i.e., the ability of one cooperator to take over the entire population,
this approach generally allows more mathematical analyses, leading to various theoretical results. Examples include the $1/3$-law \cite{Nowak2004} and the
condition $b/c>k$ for the evolution of cooperation, where $b$ is the benefit that the recipient of the cooperation gains, $c$ is the cost that a cooperator pays, and 
$k$ is the node's degree in a random regular (i.e., every node has the same degree) network
\cite{Ohtsuki06Nature}, to name a few.

An often unstated fact is that the equilibrium analysis and fixation probability analysis can yield incongruent conclusions. In fact, the effect of degree-heterogeneity of networks on evolution of cooperation is an example: degree-heterogeneous networks generally impede the fixation of cooperation relative to random regular networks \cite{Ohtsuki06Nature,Allen2017,fotouhi2019evolution}. Because experimental studies to validate theoretical and computational results on evolution of cooperation are sparse in general,
we should search for phenomena that are consistent across different theoretical frameworks. Currently, such robust understanding is lacking for a prevailing setting: evolution of cooperation in degree-heterogeneous networks.

Accumulating data verify that many empirical networks are rather temporal, i.e., time-varying (bottom row in Fig. 1) \cite{lambiotte2021guide}.
Here we theoretically investigate fixation of cooperation on temporal networks by modeling temporal networks by a sequence of time-ordered interactions [Fig. 1]. 
Although there are recent works on fixation of cooperation in temporal networks \cite{johnson2021temporal,sheng2021evolutionary,su2023strategy}, our key question is whether the degree heterogeneity of the network promotes fixation of cooperation in temporal networks.
We find that temporal degree-heterogeneous networks (in the sense that the aggregated static network is degree-heterogeneous) often favor the invasion and fixation of cooperative behavior more than temporal random regular networks. 
This result suggests that the temporal networks confer the advantages of network heterogeneity on favoring the emergence of cooperation, which is opposite to the known results for static networks~\cite{Ohtsuki06Nature,Allen2017,fotouhi2019evolution}.
Furthermore, our present results are consistent with the known results for the equilibrium dynamics: degree-heterogeneous temporal networks yield higher fractions of cooperators than degree-homogeneous temporal networks in the equilibrium, when the network is relatively large and the dynamics are initialized with a macroscopic fraction of cooperators \cite{santos2005scale}. Therefore, by viewing evolutionary dynamics of social dilemma games through lenses of temporal networks, we gain a robust understanding on the beneficial effect of the degree-heterogeneity on promotion of cooperation.

\section{Results}

\begin{figure*}[t]
	\centering
	\includegraphics[width=1 \textwidth]{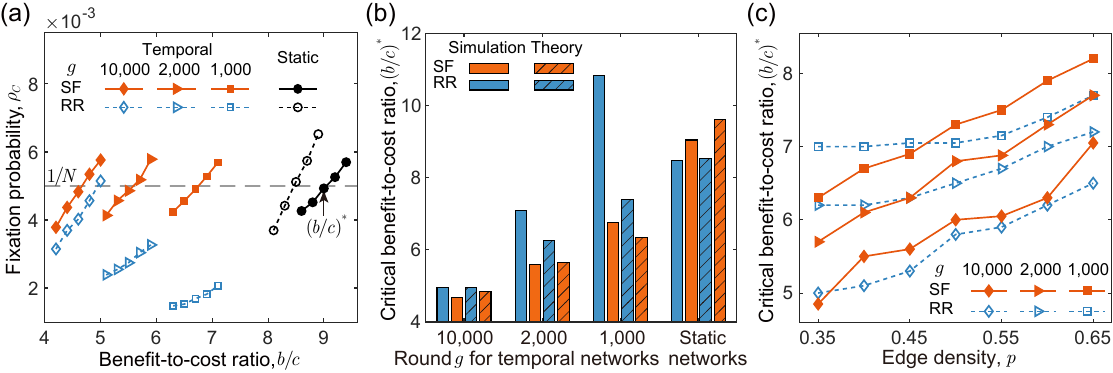}
	\caption{Comparison of the critical benefit-to-cost ratio for the fixation of cooperation between temporal scale-free (SF) networks and random regular (RR) networks.
		(a) For static networks, the critical benefit-to-cost ratio, $(b/c)^*$, above which the fixation of cooperation is favored, is higher for SF networks (indicated by the arrow pointing to the filled black dot) than the RR networks.
		For temporal SF networks (red markers), we find that $(b/c)^*$ is lower than the temporal RR networks (blue markers) over different values of the round of interactions $g$ on each snapshot.
		(b) Comparison between the numerical and theoretical results.
		(c) We further theoretically show that, when the edge density $p$ is large, $(b/c)^*$ is larger for temporal SF than that for RR networks.
		We generate synthetic temporal networks with $200$ nodes and $100$ snapshots, and the static counterparts are with the average degree of 6.
		We numerically calculate the fixation probability as the fraction of runs in which cooperators take over the whole population out of $4\times 10^5$ runs with the intensity of selection $\delta=0.01$.
	} \label{Fig2}
\end{figure*}

We consider the evolutionary game dynamics on temporal networks composed of a sequence of snapshots, where nodes indicate players and edges encode who interacts with whom [Fig. 1].
Individuals choose either cooperation ($C$) or defection ($D$).
In each round, every player $i$ interacts with its all neighbors separately and accumulates the obtained payoffs.
At the end of each round, an individual is randomly chosen for updating its strategy \cite{perc2013evolutionary,Ohtsuki06Nature}, where the individual either imitates the strategy of its neighbor $j$ with the probability proportional to the fitness of $j$, denoted by $F_j$, or retains its strategy with the probability proportional to its own fitness.
We set  $F_j = 1 + \delta f_j$, where $f_j$ is the accumulated payoff for player $j$ in the current round, and $0<\delta\ll 1$ specifies the weak intensity of selection \cite{Allen2017}.
Starting with a single cooperator randomly placed in a population with $N-1$ defectors, the evolutionary process involves $g$ rounds of interactions on each snapshot before switching to the next snapshot. Then, strategy update happens at the end of each round. The process ends when all players become either cooperator or defector [Fig. S1] \cite{li2020evolution}.
The fixation probability of cooperation ($\rho_C$) on temporal networks is defined by the probability that a single cooperator takes over the entire population \cite{Nowak2004,Ohtsuki06Nature,Allen2017}.

Static random regular networks are known to present a lower critical benefit-to-cost ratio $(b/c)^*$ than static degree-heterogeneous networks (e.g., scale-free networks, which are defined by a power-law degree distribution), where $(b/c)^*$ is the threshold above which natural selection favors the invasion and replacement of cooperation in a population full of defectors, namely, $\rho_C>1/N$ \cite{Ohtsuki06Nature,Allen2017}.
Note that $1/N$ is the fixation probability in the case of a neutral drift, i.e., when the invader has the same fitness as the resident, on both static and temporal networks [Fig. S2].
Here we first explore the fixation of cooperation on temporal scale-free and random regular networks compared to the corresponding static networks.
After numerically verifying the above result in Fig. 2a (shown by the filled and empty circles), we further show that temporal scale-free networks promote the fixation of cooperation (i.e., lower $(b/c)^*$) more than their static counterparts.

Surprisingly, we find that the \textit{temporal} scale-free networks facilitate the fixation of cooperation more than the temporal random regular networks by yielding a lower $(b/c)^*$ at different values of $g$.
This result is in sharp contrast with the previous results comparing \textit{static} scale-free and random regular networks \cite{Ohtsuki06Nature,Allen2017,fotouhi2019evolution}.
We generate each snapshot in synthetic temporal networks by randomly activating a fraction $p$ of edges in the underlying static networks;
we set $p=0.3$ in Fig. 2a.

\begin{figure*}
	\centering
	\includegraphics[width= 1 \textwidth]{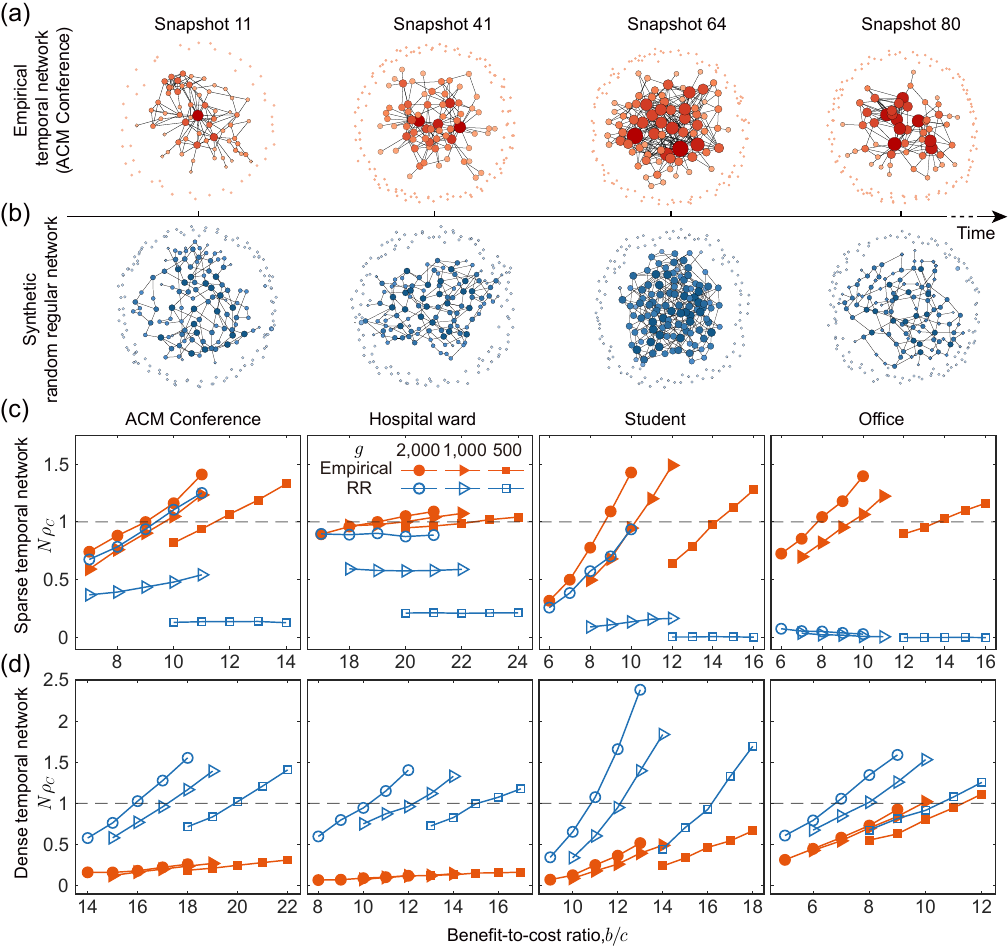}
	\caption{Fixation probability of cooperation on empirical temporal networks and the corresponding synthetic random regular networks.
	(a) Construction of empirical temporal networks based on the dataset capturing face-to-face proximity of $113$ conference attendees over about $2.5$ days in the ACM conference.
	We obtain the sequence of snapshots by aggregating interactions during the time interval $\Delta t=30~\text{mins}$.
	(b) Snapshots in synthetic temporal random regular networks are generated at the same number of edges as that of the corresponding snapshot in the empirical temporal networks.
	(c) For four empirical datasets, i.e., ACM conference, Hospital ward, Student and Office, we numerically show that sparse empirical temporal networks ($\Delta t=30~\text{mins},30~\text{mins},2~\text{hours},2~\text{hours}$) yield a higher fixation probability (red markers) than the corresponding temporal random regular (RR) networks (blue markers), while the opposite holds true for dense empirical temporal networks ($\Delta t=6~\text{hours},6~\text{hours},12~\text{hours},12~\text{hours}$) as shown in (d).
	The details of the empirical dataset are presented in table S1 and Figs. S7--S10.
	} \label{Fig3}
\end{figure*}
\begin{figure*}[t]
	\centering
	\includegraphics[width=1 \textwidth]{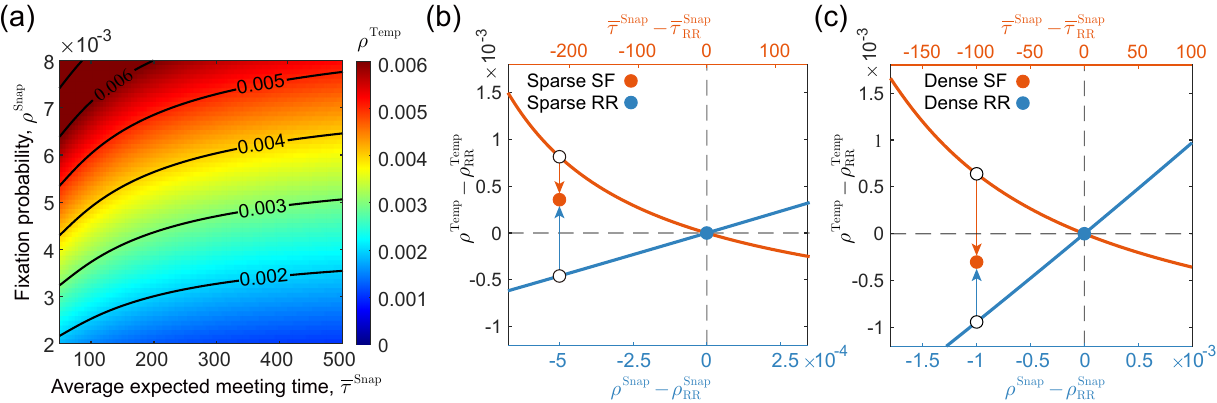}
	\caption{The relation between the fixation probability and time on each snapshot determines the role of temporal scale-free networks on favoring cooperation.
	(a) We theoretically show the contour of the fixation probability of cooperation on temporal networks ($\rho^{\text{Temp}}$) as a function of the fixation probability ($\rho^{\text{Snap}}$) and fixation time ($\overline{\tau}^{\text{Snap}}$) on each snapshot.
	(b) By setting the fixation probability ($\rho^{\text{Snap}}_{\text{RR}}$) and fixation time ($\overline{\tau}^{\text{Snap}}_{\text{RR}}$) on the snapshot of temporal random regular (RR) networks as the baseline,
	we present how the alteration of $\rho^{\text{Snap}}$ and $\overline{\tau}^{\text{Snap}}$ changes the fixation probability $\rho^{\text{Temp}}$ on a temporal network.
	Specifically, relative to the case for a sparse temporal RR network (blue solid dot, $\rho^{\text{Temp}}=\rho^{\text{Temp}}_{\text{RR}}$), a sparse temporal SF network (red solid dot, which is obtained by the summation of two white solid dots) yields the change in $\rho^{\text{Temp}}$ caused by $\overline{\tau}^{\text{Snap}}-\overline{\tau}^{\text{Snap}}_{\text{RR}}$ (red line) being larger than the change caused by $\rho^{\text{Snap}}-\rho^{\text{Snap}}_{\text{RR}}$  (blue line).
	This combined effect increases the corresponding fixation probability on sparse temporal SF networks.
	(c) In dense networks, the increase in $\rho^{\text{Temp}}$ caused by $\overline{\tau}^{\text{Snap}}-\overline{\tau}^{\text{Snap}}_{\text{RR}}$ (red line) does not compensate the decrease in $\rho^{\text{Temp}}$ caused by $\rho^{\text{Snap}}-\rho^{\text{Snap}}_{\text{RR}}$ (blue line), leading to the smaller fixation probability for temporal SF than RR networks (red solid dot).
	We construct the temporal networks in the same manner as that for Fig. 2 and set $g=10^4$.
	We also set $p=0.3$ in (b) and $p=0.7$ in (c).
	} \label{Fig4}
\end{figure*}

To theoretically investigate the fixation of cooperation on temporal networks, we develop an analytical framework. We denote the probability of having $n$ cooperators at the beginning of the $m$th snapshot by $p_m(n)$, and describe the state of the system by a probability vector $\boldsymbol p_m=(p_m(0),p_m(1),\ldots,p_m(N))^{\top}$ with $\sum_{n=0}^N p_m(n)=1$, where ${}^{\top}$ represents the transposition.
Since each snapshot is dominated by a largest connected component [Fig. S3],
we consider the evolutionary dynamics on the largest connected component with $N_m$ players of the $m$th snapshot.

We denote by $\mathcal{T}^{g}_{m}(l, h)$ the transition probability starting from $l$ cooperators in the largest connected component in the $m$th snapshot and ending with $h$ cooperators after $g$ rounds of evolution.
To approximate $\mathcal{T}^{g}_{m}(l, h)$, we consider the weighted average frequency of cooperators at the $g$th round, denoted by $\mathbb{E}_{\mathbf{u}}\left[ \hat{x}(g)\right]$. 
Specifically, we have $\hat{x}(g)=\sum_i \tilde{\pi}_i x_i(g)$, where $x_i(g)\in \{0,1\}$ represents the strategy of node $i$ at the $g$th round (i.e., either cooperation ($1$) or defection ($0$));
$\tilde{\pi}_{i}=(\sum_j w_{ij} +1)/(\sum_{i,j} w_{ij} +N_m)$ is the reproductive rate of node $i$;
$w_i=\sum_{i,j} w_{ij}$ is the weighted degree for node $i$; 
$w_{ij}$ indicates the edge weight between $i$ and $j$ (see the Supplementary Materials).
Based on the evolutionary dynamics, we have
\begin{equation}  
	\begin{aligned}
		\mathbb{E}_{\mathbf{u}}\left[ \hat{x}(g)\right]
		&=\frac{l}{N_m}+\frac{l(N_m-l)\delta }{2N_m(N_m-1)}\left[-c \sum\limits_{i,j}\tilde{\pi}_{i}\tilde{p}_{ij}^{(2)}w_j \tilde{\tau}_{ij}\left(T\right)\right.\\
		&+b \left. \left(\sum\limits_{i,j,k}\tilde{\pi}_{i}\tilde{p}_{ij}^{(2)}w_j p_{jk} \tilde{\tau}_{ik}\left(T\right)-\sum\limits_{i,j}\tilde{\pi}_{i} w_i \tilde{p}_{ij} \tilde{\tau}_{ij}\left(T\right)  \right)\right],~~~~~
	\label{mean_freq}\nonumber
	\end{aligned}
\end{equation}
where $p_{ij}^{(n)}$ is the transition probability of an $n$-step random walk from $i$ to $j$;
$\tilde{p}_{ij}^{(n)}$ is the probability that $i$ imitates the strategy of $j$ after an $n$-step random walk during the spreading of $j$'s strategy;
$\mathrm{P}_{\{i, j\}}(\tau )$ represents the probability that two random walkers (one starting at node $i$ and the other at node $j$) meet at time $\tau$ (see the Supplementary Materials); $\tilde{\tau}_{i j}(T)=\sum_{t=0}^{T} \sum_{\tau=t+1}^{\infty} \mathrm{P}_{\{i, j\}}(\tau )$ represents the accumulation of the probability that $i$ and $j$ have not met until $t$.
Here $T\equiv \lfloor 2(g-1)/N_m \rfloor$ indicates the effective $g$ that is rescaled by the probability $2/N_m$ with which the individual chosen for updating is from the two random walkers (see the Supplementary Materials).
Based on the probability distribution of the number of cooperators starting from $l$ cooperators through $g$ rounds of interaction we observed numerically in Fig. S4, we theoretically write the transition probability as
\begin{equation}
	\begin{aligned}
		&\mathcal{T}^{g}_{m}(l, 0) \approx \left(1-\lim\limits_{g'\to\infty}\mathbb{E}_{\mathbf{u}}\left[ \hat{x}(g')\right]\right) \left[1- \mathrm{exp}\left(\frac{2(1-N_m) g}{lN_m\tau_{m}}\right)\right], \\  \nonumber
		&\mathcal{T}^{g}_{m}(l, N_m) \approx  \lim\limits_{g'\to\infty}\mathbb{E}_{\mathbf{u}}\left[ \hat{x}(g')\right]\left[1-\mathrm{exp}\left(\frac{2(1-N_m) g}{(N_m-l)N_m\tau_{m}}\right)\right],                       \\ 
		&\mathcal{T}^{g}_{m}(l, h)  \approx \alpha \beta^{|h-l|} \quad  (0<h<N_m),
		\label{trans_approx}\nonumber
	\end{aligned}
\end{equation}
where $\tau_m$ is the expectation of the meeting time averaged over the component of $N_m$ players in the snapshot $m$.
When $l=1$, $\lim_{g'\to\infty}\mathbb{E}_{\mathbf{u}}\left[ \hat{x}(g')\right]$ is equivalent to the fixation probability of cooperation on the component.
Based on the definition of the transition probability, we obtain the above parameters $\alpha$ and $\beta$ by imposing
\begin{equation}
	\sum_{h=0}^{N_m} \mathcal{T}^{g}_{m}(l, h)=1
\end{equation}
and
\begin{equation}
	\sum_{h=0}^{N_m} \frac{h}{N_m}\mathcal{T}^{g}_{m}(l, h)\approx\mathbb{E}_{\mathbf{u}}\left[ \hat{x}(g)\right].
	\nonumber
\end{equation}

Under the approximation that cooperators are uniformly distributed at the onset of each snapshot,
the probability that there are $l$ cooperators in the largest connected component of $N_m$ players is $q_m(n, l)=\tbinom{n}{l}\tbinom{N-n}{N_m-l}/\tbinom{N}{N_m}.$
Therefore, the state of the system between two successive snapshots $m$ and $m+1$ obeys the master equation 
\begin{equation}
	p_{m+1}(s)=\sum_{n-l+h=s} p_m(n) q_m(n, l) \mathcal{T}^{g}_{m}(l, h).
	\label{trans_prob}
\end{equation}
By definition, the fixation probability of cooperation on temporal networks is
\begin{equation}
	\rho_C =\lim_{m\to\infty} p_{m}(N).
	\label{eq:rho_C-final}
\end{equation}
We corroborate our numerical findings with these theoretical results in Fig. 2b.
We obtained $(b/c)^*$ as the value of $(b/c)$ at which $\rho_C$ given by Eq.~\eqref{eq:rho_C-final} is equal to $1/N$.

For synthetic temporal networks, with the increase in the edge density, $p$, the snapshot becomes dense, and the network tends to be static, which diminishes the effect of network temporality on favoring cooperation. For example, at $p = 0.7$, similar to the static scenario, the temporal random regular network performs better than the temporal scale-free network at fostering cooperation both theoretically and numerically [Fig. S5].
We theoretically find that temporal scale-free networks yield a smaller value of $(b/c)^*$ (i.e., easier cooperation) than temporal random regular networks when $p < p^*$ and vice versa when $p > p^*$, where $p^*$ decreases as $g$ increases [Fig. 2c].

Moreover, for empirical temporal networks which are constructed from empirical social interactions with different widths of time window $\Delta t$ [Fig. 3a], we confirm our findings. 
Specifically, we find that, for different values of $g$, cooperation fixates with a higher probability for empirical temporal networks [Fig. 3a] than the corresponding temporal random regular networks [Fig. 3b] when the snapshots are sparse [Fig. 3c].
The opposite is the case when the snapshots are dense, which we realize by increasing $\Delta t$ [Fig. 3d].
These results are consistent with the numerical and theoretical findings on synthetic temporal networks shown in Fig. 2.

To intuitively understand the transition point in terms of the edge density we observed in both synthetic and empirical temporal networks,
we analyze the fixation probability of cooperation on temporal networks at $g=10^4$, with which the fixation of full cooperators or defectors tends to be reached within the largest connected component of each snapshot.
In this case, the probability of the transient state $\mathcal{T}^{g}_{m}(l, h)~(0<h<N_m)$ can be represented by $\left[1-\mathcal{T}^{g}_{m}(l, 0)-\mathcal{T}^{g}_{m}(l, N_m)\right]/(N_m-1)$ according to Eq.~\eqref{trans_approx}.
For ease of understanding, we now consider that each snapshot is connected and presents the same fixation probability $\rho_C^{\text{Snap}}$ and the average expected meeting time $\overline{\tau}^{\text{Snap}}$.
The probability distribution of the number of cooperators on temporal networks as $m$ tends to infinity is
\begin{equation}
	\lim_{m\to\infty}\boldsymbol p_m^{\top}=\boldsymbol p_{0}^{\top}\prod_{m=0}^{\infty} \mathcal{T}^{g}_{m}, \label{eq4}
\end{equation}
where $\mathcal{T}^{g}_{m}$ is the transition probability on each snapshot, which is fully determined by $\rho^{\text{Snap}}$ and $\overline{\tau}^{\text{Snap}}$.
Based on Eq.~\eqref{eq4}, we demonstrate that a shorter $\overline{\tau}^{\text{Snap}}$ may sustain the same fixation probability of cooperation on temporal networks (i.e., $\rho^{\text{Temp}}$) at a lower fixation probability on each snapshot (i.e., $\rho^{\text{Snap}}$) [Fig. 4a].
To further understand the role of $\overline{\tau}^{\text{Snap}}$ in temporal evolutionary dynamics, we calculate the average time $\langle t_C \rangle$ that a single cooperator requires to take over the entire population on each snapshot, namely, i.e., the fixation time.
Using Eq.~\eqref{trans_approx}, we obtain
\begin{equation}
	\langle t_C \rangle = \frac{1}{\rho_C^{\text{Snap}}} \sum_{g=1}^{\infty} g \left(\mathcal{T}^{g}_{m}(1,N)-\mathcal{T}^{g-1}_{m}(1,N)\right) \approx \frac{N \overline{\tau}^{\text{Snap}}}{2}. \nonumber
\end{equation}
In other words, the average expected meeting time $\overline{\tau}^{\text{Snap}}$ is proportional to the fixation time of cooperation on each snapshot $\langle t_C \rangle$.
This relationship actually uncovers why temporal scale-free networks favor the fixation of cooperation more than the temporal random regular networks in the case of sparse snapshots.
Specifically, at the same $b/c$, although the fixation probability for static scale-free networks is lower than for static random regular networks, the former has a shorter fixation time, which facilitates the emergence of cooperative clusters in each snapshot at finite rounds of interactions, $g$.
On the contrary, the static random regular networks require a long time for fixation, which makes it more difficult for cooperators to diffuse and cluster at finite $g$.
Therefore, sparse temporal scale-free networks can achieve a larger fixation probability than sparse temporal random regular networks by taking advantage of shorter fixation time on each snapshot (red dot in Fig. 4b).
As snapshots become denser, the difference in the fixation time between random regular and scale-free snapshots shrinks.
Then, the disadvantage from the smaller fixation probability for the static scale-free networks can not be compensated with its shorter fixation time [Fig. 4c].
This trade-off between the fixation probability and fixation time is also confirmed in empirical temporal networks [Figs.~3c and 3d].

Finally, we explore effects of the round of interactions on each snapshot, $g$.
In general, a larger $g$ allows the emergence and preservation of clusters of cooperators before the switching of snapshots, resulting in a higher fixation probability for cooperation, as shown in Figs. 2 and 3.
The fixation probability on temporal networks in the limit $g \to\infty$ is determined solely by the fixation probability on the first snapshot, i.e., a single static network.
This indicates that the advantages of temporal scale-free networks on favoring cooperation brought by the shorter fixation time are discounted as $g$ increases,
which are observed in both synthetic and empirical temporal networks [Figs. 2 and 3].
When the network structure changes more rapidly than the dynamics occurring on the network (i.e., $g<1$), our main results still hold qualitatively [Fig. S6].

\section{Discussion}

To sum up, we have found that temporal scale-free networks may provide more benefits for the fixation of cooperation than temporal random regular networks, contrary to previous results reported for the static networks.
The trade-off between the fixation probability and fixation time that we have discovered explains the advantage of temporal scale-free networks and also the emergence of the turning point as the network structures tend to be static, which we further verified on empirical temporal networks.
The present theoretical and numerical results enable us to draw a unified conclusion on the effect of degree heterogeneity on evolution of cooperation.
Specifically, while static degree-heterogeneous networks promote cooperation under the equilibrium dynamics, the same networks suppress cooperation under non-equilibrium fixation dynamics. In contrast, temporal degree-heterogeneous networks promote cooperation under both types of dynamics. The equilibrium and non-equilibrium approaches have complementary strengths. We encourage that both existing and future results on evolutionary dynamics are corroborated by both approaches.

By numerically studying the equilibrium frequency of cooperators, our previous work~\cite{cardillo2014evolutionary,li2020evolution} reported that \textit{temporal} networks facilitate the emergence of cooperation compared to their \textit{static} counterparts.
Such advantages of temporal networks in promoting cooperation have also been supported by studies of fixation probability of cooperation in temporal networks \cite{sheng2021evolutionary,su2023strategy}.
Our present results also support that temporal networks facilitate cooperation compared to static networks under a wide range conditions. Therefore, the promotion of cooperation in temporal networks compared to their static network counterparts seems to be another phenomenon commonly observed for the equilibrium and non-equilibrium evolutionary game dynamics. However,
because models of temporal networks employed vary across these studies, this topic warrants for further work.

We have shown that the advantage of temporal scale-free networks disappears as the edge density (i.e., fraction of activated edges) of the network increases.
We have theoretically and numerically shown that the superiority of temporal scale-free networks originates from short fixation time of cooperation on snapshot networks, which paves the way for the emergence of clusters of cooperators before the network switches to a next one.
There is a large body of research analyzing the fixation probability or fixation time on static network structures \cite{Ohtsuki06Nature,Allen2017,Altrock2009,Su2019,mcavoy2021fixation,Zhou2015,tkadlec2019population}, yet how they couple to affect the temporal evolutionary dynamics remains unknown.
Our work opens the door to exploring evolutionary dynamics on temporal networks through those key properties.

Considering that different individuals may interact with their neighbors at diverse rates and rhythms, a promising direction for future research is the design of temporal interaction structures and mechanisms to boost cooperation in a given population. For example, because the synchronization speed in temporal networks considerably depends on the order of edges to be sequentially used \cite{masuda2016accelerating}, the ease of cooperation (e.g., $(b/c)^*$) may be as well.
Moreover, group interactions that capture collective interactions with multiplayer games \cite{hilbe2014cooperation} on underlying exogenous dynamic structures or even higher-order networks \cite{majhi2022dynamics} may lead to more exotic evolutionary dynamics.
Our findings---that temporal heterogeneous networks facilitate the emergence of cooperation---pave the way for future investigations on temporal networks underlying realistic complex systems.

\end{document}